\documentclass[prd,nofootinbib,english]{revtex4}
\usepackage{amsmath}
\usepackage{amsfonts,color}
\usepackage{amssymb,float}
\usepackage[utf8]{inputenc}
\usepackage{color}
\usepackage[unicode=true,pdfusetitle,
 bookmarks=true,bookmarksnumbered=true,bookmarksopen=true,bookmarksopenlevel=3,
 breaklinks=false,pdfborder={0 0 1},backref=false,colorlinks=true]
 {hyperref}
\usepackage{enumitem}

\usepackage{tikz}
\usetikzlibrary{shapes.geometric}
\usetikzlibrary{arrows.meta,arrows}

\newcommand{\udt}[3]{#1^{#2}_{\phantom{#2}#3}}
\newcommand{\udut}[4]{#1^{#2\phantom{#3}#4}_{\phantom{#2}#3\phantom{#4}}}

\newcommand{\dut}[3]{#1_{#2}^{\phantom{#2}#3}}
\newcommand{\dudt}[4]{#1_{#2\phantom{#3}#4}^{\phantom{#2}#3}}

\begin{document}
\title{Can Horndeski Theory be recast using Teleparallel Gravity?}

\author{Sebastian Bahamonde}
\email{sbahamonde@ut.ee, sebastian.beltran.14@ucl.ac.uk}
\affiliation{Laboratory of Theoretical Physics, Institute of Physics, University of Tartu, West Ostwaldi 1, 50411 Tartu, Estonia}
\affiliation{Department of Mathematics, University College London, Gower Street, London, WC1E 6BT, United Kingdom}

\author{Konstantinos F.	Dialektopoulos}
\email{dialektopoulos@na.infn.it}
\affiliation{Center for Gravitation and Cosmology, College of Physical Science and Technology, Yangzhou University, Yangzhou 225009, China,}
\affiliation{Aristotle University of Thessaloniki, Thessaloniki 54124, Greece}
\affiliation{Institute of Space Sciences and Astronomy, University of Malta, Msida, MSD 2080, Malta}
	
\author{Jackson Levi Said}
\email{jackson.said@um.edu.mt}
\affiliation{Institute of Space Sciences and Astronomy, University of Malta, Msida, MSD 2080, Malta}
\affiliation{Department of Physics, University of Malta, Msida, MSD 2080, Malta}

\begin{abstract}
Horndeski gravity is the most general scalar tensor theory, with a single scalar field, leading to second-order field equations and after the GW170817 it has been severely constrained. Since this theory is very important in modified gravity, it is then worth studying possible similar theories starting from other frameworks. In this paper, we study the analog of Horndeski's theory in the Teleparallel Gravity framework where gravity is mediated through torsion instead of curvature. We show that, even though, many terms are the same as in the curvature case, we have much richer phenomenology in the teleparallel setting because of the nature of the torsion tensor. Moreover, teleparallel Horndeski contains the standard Horndeski gravity as a subcase and also contains many modified Teleparallel theories considered in the past, such as $f(T)$ gravity or teleparallel dark energy. Thus, due to the appearance of a new term in the Lagrangian, this theory can explain dark energy without a cosmological constant, may describe a crossing of the phantom barrier, explain inflation and also solve the tension for $H_0$, making it a good candidate for a correct modified theory of gravity.
\end{abstract}

\maketitle

\section{Introduction}
In spite of their great success, $\Lambda$CDM and General Relativity (GR) are plagued with many shortcomings. The value of the cosmological constant~\cite{RevModPhys.61.1}, the nature of dark matter and dark energy, the nature of singularities and also its inability to provide a quantum description of gravity are only some of them. Followed by the will to frame these shortcomings in a self-consistent cosmological model, and in general, in a gravity theory that ``works'' at all scales, scientists started to pursue modifications to the standard cosmological model. One of the first and simplest modifications was suggested by Brans and Dicke in 1961 \cite{Brans:1961sx}. They introduced a new scalar field, nonminimally coupled to the Ricci scalar that effectively played the role of a varying Newton's constant. The literature on the theory is exhausting and the interested reader is referred to reviews on modified gravity \cite{Clifton:2011jh,Capozziello:2011et,Nojiri:2017ncd}. 

A decade later, in the beginning of the 1970s, Horndeski wrote down the most general scalar-tensor theory, with a single scalar field that leads to second-order field equations \cite{1974IJTP...10..363H}. However, it did not receive much attention until the late 2000s, when it was realized that all the terms in Horndeski's theory originate from Galileons \cite{Nicolis:2008in}. Finally, its current form, in curved spacetime, was given by Deffayet, Deser, and Esposito-Farese \cite{Deffayet:2009mn}. 

Many known modifications of gravity, from Brans-Dicke theory, k-essense and kinetic braiding to the scalar representation of $f(\mathring{R})$ gravity (we will explain the notation later in this section) can be considered as subcases of Horndeski gravity. A lot of ink has been spilled in studies of Horndeski gravity and one can see the reviews \cite{Kobayashi:2019hrl,Deffayet:2013lga} for more details. More in detail, black hole solutions have been found in \cite{Babichev:2016rlq,Babichev:2015rva,Babichev:2017guv}, neutron stars in \cite{Maselli:2016gxk,Cisterna:2015yla}, and inflation has been studied in \cite{Kobayashi:2011nu}. In addition, self-accelerating solutions are presented in \cite{Silva:2009km} and the Vainshtein mechanism, that is a screening mechanism to ``hide'' the scalar field at small scales, is discussed among others in \cite{Khoury:2010xi}.

It is known today that there exists an alternative formulation of gravity, the so-called \textit{Teleparallel equivalent of General Relativity} (TEGR) or Teleparallel Gravity \cite{Aldrovandi:2013wha}, that is completely equivalent to GR at the level of the field equations, and instead of curvature, it uses torsion to describe the gravitational interactions. This theory uses the tetrad formalism\footnote{It is remarkable to note that Einstein himself wanted to incorporate General Relativity in a unified theory with electrodynamics and for this reason he used the tetrad formalism. His attempt was not successful, because he thought that the extra degrees of freedom of the tetrad could play somehow the role of the electromagnetic field, but this was not the case.}, meaning that the dynamical field is not the metric anymore, but rather a tetrad field defined on a tangent space at each point of the general manifold. In this framework, gravity is no longer the effect of geometry of the spacetime but rather a force, just like the Lorentz force in electrodynamics. 

One could reasonably ask, why should one study a different theory of gravity if at the level of the equations of motion it is equivalent to GR? As it turns out, Teleparallel Gravity has several features which may be more natural when compared with GR. First of all, it is a gauge theory of translations, meaning that it can be more easily unified with the three other fundamental forces of the Standard Model~\cite{aldrovandi1995introduction}. As a gauge theory, it could even survive in the absence of the equivalence principle~\cite{Aldrovandi:2003pa}. Moreover, in the framework of TEGR one can separate gravitational from inertial effects (in the correct gauge), and because of that, one can define a gravitational energy-momentum density~\cite{deAndrade:2000kr}.

As in the curvature case, however, one cannot explain all the observations with pure TEGR; thus, one should look for modifications. In Ref.~\cite{Cai:2015emx}, a review on $f(T)$ theories is presented. Scalar fields have also been considered in numerous ways with some indicative being \cite{Chakrabarti:2017moe,Said:2017nti,Bahamonde:2018miw,Geng:2011ka,Kofinas:2015hla,Xu:2012jf,Geng:2011aj,Bahamonde:2015hza,Bahamonde:2016jqq,Hohmann:2018rwf,Bahamonde:2019gjk,Zubair:2016uhx}. In the recent series of papers \cite{Hohmann:2018vle,Hohmann:2018dqh,Hohmann:2018ijr}, a serious attempt to construct a general scalar-torsion theory has been made. Finally, during the preparation of this work, this paper appeared \cite{Gonzalez-Espinoza:2019ajd} where the authors consider a subcase of the theory presented here and they study inflation.  

After the recent discovery of gravitational waves and specifically after the electromagnetic counterpart of GW170817, most of Horndeski's terms are severely constrained by the tensor mode propagation speed \cite{Kreisch:2017uet,Gong:2017kim}. That is the reason why we wanted to study the Horndeski analog in the teleparallel framework, in order to see if the constrained terms in the curvature case could survive in this setting. As we will see, this indeed might be the case.

What we do in this paper is to construct the most general scalar-torsion theory, with a single scalar field, that 1) leads to second-order field equations for the tetrad (or equivalently to the metric) and the scalar field, 2) it is not parity-violating and 3) contains at most quadratic terms of the torsion tensor. As we will see later in the paper, this theory has a lot more phenomenology than in the curvature case because there appear many new scalars, and in addition, the higher-order derivative couplings ($\mathring{L}_4$ and $\mathring{L}_5$) could in principle survive the gravitational wave analysis. 

The paper is organized as follows: In the next section \S.\ref{sec:Intro_Tele_Grav}, we introduce Teleparallel Gravity and present the irreducible decomposition of the torsion tensor. At the end of this section, we also show how the covariantization procedure occurs, i.e. what is the gauge structure of the theory. Furthermore, in \S.\ref{sec:HorndeskiTeleparallel}, we present the conditions on which we build our theory, we construct all the possible scalars and we present the teleparallel Horndeski theory. Finally, in \S.\ref{sec:CosmoTeleDeski}, we consider a cosmological (flat Friedmann–Lema\^{i}tre–Robertson–Walker (FLRW)) background and after calculating explicitly all the new scalars that appear in the theory, we write down the equation of motion for the scalar field and the scale factor. We notice that standard Horndeski gravity is a subclass of teleparallel Horndeski gravity since new extra terms appear in the final form of the action of the theory.

Throughout the paper the Latin indices $i,j,k,...$ represent coordinates on the tangent space, while the Greek indices $\alpha,\beta,\mu,...$ represent coordinates of the general manifold. Quantities calculated with the Levi-Civita connection (e.g. connections, covariant derivatives, d'Alembertians) are given with a circle on top, e.g. $\mathring{\nabla}_{\mu}$ and quantities referring to flat spacetime are denoted with a bar on top, e.g. $\bar{\Box}$. All the other quantities that have no symbols, e.g. $\Gamma ^{\alpha}{}_{\mu\nu}$, are calculated with (or referred to) the Weitzenb\"ock connection. Also unless otherwise stated, we use the metric signature $\eta_{\mu\nu}=\textrm{diag}(-1,1,1,1)$, and geometric units.

\section{Teleparallel Gravity}

\subsection{The Teleparallel Equivalent of General Relativity and its Decomposition}
\label{sec:Intro_Tele_Grav}

In GR, gravity is expressed through curvature by means of the Levi-Civita connection, $\mathring{\Gamma}{}^{\sigma}{}_{\mu\nu}$, in the context of Riemann geometry. However, geometric deformations can be characterized using other geometric quantities, or connections. In fact, there exists a trinity of characterizations of gravity such that GR can be reproduced at the level of the field equations \cite{BeltranJimenez:2019tjy}. In this work, we consider the setting of Teleparallel Gravity \cite{aldrovandi1995introduction,Cai:2015emx,Krssak:2018ywd} which carries a fundamental distinction from curvature-based descriptions of gravity. Here, the Levi-Civita connection is replaced by the Weitzenb\"{o}ck connection, $\Gamma^{\sigma}{}_{\mu\nu}$, which is curvatureless while still observing the metricity condition, and is given by
\begin{equation}\label{eq:weitzenbockdef}
\udt{\Gamma}{\sigma}{\mu\nu} := \dut{h}{a}{\sigma}\partial_\mu \udt{h}{a}{\nu} + \dut{h}{a}{\sigma}\udt{\omega}{a}{b\mu}\udt{h}{b}{\nu}\,,
\end{equation}
where $\udt{h}{a}{\rho}$ is the tetrad field, and $\udt{\omega}{a}{b\mu}$ the spin connection. This is the most general linear affine connection that is both curvatureless and satisfies the metricity condition \cite{aldrovandi1995introduction}. The tetrad represents transformations between the general manifold and the tangent space, and $\dut{h}{a}{\mu}$ represents the tetrad inverse. This property of connecting tangent space and manifold tensors is called soldering, and is a very useful tool for raising inertial relations to the general manifold. For instance, consider the inertial four-momentum $P^a$ which can be raised to $P^{\mu}=\dut{h}{a}{\mu}P^a$ for the general manifold. \\

Given the intrinsic link between inertial and noninertial indices in Teleparallel Gravity, it follows that some caution needs to be taken when constructing a theory that is invariant under local Lorentz transformations (LLTs). Any gravitational theory should be invariant under LLTs. This is the role that the spin connection plays. To account for this invariance, Teleparallel Gravity incorporates so-called spin connections, $\udt{\omega}{a}{b\mu}$, which sustains this freedom. GR also features spin connections but these are mainly hidden in the inertial structure of the theory \cite{aldrovandi1995introduction}. Together the tetrad and spin connection specify the frame analogous to the metric tensor scenario in GR. Thus, the spin connection is not a second degree of freedom of the gravitational component of the theory, but a regular object used in any theory invariant under LLTs. \\

Given the full breath of LLTs (Lorentz boosts and rotations), the tetrad can be transformed by its inertial index through
\begin{equation}
h'{}^{a}{}_\mu=\udt{\Lambda}{a}{b}\udt{h}{b}{\mu}\,,
\end{equation}
where $\udt{\Lambda}{a}{b}$ is a LLT. In this way, the spin connection can also be represented as the combination of completely inertial LLTs in the form \cite{Krssak:2015oua}
\begin{equation}
\udt{\omega}{a}{b\mu} = \udt{\Lambda}{a}{c}\partial_{\mu}\dut{\Lambda}{b}{c}\,,
\end{equation}
which preserves the LLT invariance of the theory as a whole. \\

On the other hand, the metric tensor $g_{\mu\nu}$ characterizes the general manifold through distance measurements, while the tetrad, $\udt{h}{a}{\mu}$ relates the tangent space with the manifold. For consistency, they also observe the relations
\begin{align}
\udt{h}{a}{\mu}\dut{h}{b}{\mu} &= \delta^a_b\,,\\
\udt{h}{a}{\mu}\dut{h}{a}{\nu} &= \delta^{\nu}_{\mu}\,,
\end{align}
which form the orthogonality conditions of the setup. More generally, since the tetrad fields raise inertial indices, they can be used to relate the Minkowski and general manifold metric tensors through the equations
\begin{align}
g_{\mu\nu} &= \udt{h}{a}{\mu}\udt{h}{b}{\nu}\eta_{ab}\,,\label{Min_trans_Gen}\\
\eta_{ab} &= \dut{h}{a}{\mu}\dut{h}{b}{\nu}g_{\mu\nu}\,,
\end{align}
where the active role of the tetrad can be viewed as a field that replaces the metric as the fundamental dynamical variable of the theory. Here, the position dependence of these relations has been suppressed for brevity's sake.\\

At this point, we need to make a distinction here between two kinds of tetrads. Firstly, trivial tetrads are those tetrad fields that represent manifolds that are nongravitational, and so are effected only by LLTs. In terms of Eq.~(\ref{Min_trans_Gen}), this then takes the form $\eta_{\mu\nu} = \udt{e}{a}{\mu}\udt{e}{b}{\nu}\eta_{ab}$, where $\udt{e}{a}{\mu}$ symbolizes the use of trivial tetrad fields. Alternatively, gravitational systems make use of nontrivial tetrad fields represented by $\udt{h}{a}{\mu}$. \\

In this construction, the curvature measured by the Riemann tensor will always be zero, while the torsion will depend on the form of the tetrad and spin connection components. This is not to say that the Riemann tensor computed with the Levi-Civita connections is zero, but that the curvature of the theory is zero. Torsion can then be characterized as an antisymmetric property through \cite{Bahamonde:2017wwk}
\begin{equation}
\udt{T}{a}{\mu\nu} :=  2\Gamma^{a}{}_{[\mu\nu]}\,,
\end{equation}
which is a measure of the field strength, and where square brackets represent the antisymmetric operator ($A_{[\mu\nu]}=\frac{1}{2}\left(A_{\mu\nu}-A_{\nu\mu}\right)$). $\udt{T}{a}{\mu\nu}$ is called the torsion tensor and transforms covariantly under both diffeomorphisms and LLTs. \\

The torsion tensor is the fundamental measure of torsion, analogous to the Riemann tensor for curvature. However, we can define other useful tensors, such as the contorsion tensor which is the difference between the Levi-Civita $\mathring{\Gamma}{}^{\sigma}{}_{\mu\nu}$ and Weitzenb\"{o}ck connections $\Gamma^{\sigma}{}_{\mu\nu}$~\cite{Cai:2015emx,RevModPhys.48.393}
\begin{equation}\label{contorsion_def}
\udt{K}{\sigma}{\mu\nu} := \Gamma^{\sigma}{}_{\mu\nu} - \mathring{\Gamma}{}^{\sigma}{}_{\mu\nu} = \frac{1}{2}\left(\dudt{T}{\mu}{\sigma}{\nu}+\dudt{T}{\nu}{\sigma}{\mu}-\udt{T}{\sigma}{\mu\nu}\right),
\end{equation}
which plays an important role in relating Teleparallel Gravity with Levi-Civita based theories. Another important ingredient in forming a Teleparallel theory of gravity is the so-called superpotential which is defined as
\begin{equation}
\dut{S}{a}{\mu\nu} := \udt{K}{\mu\nu}{a} - \dut{h}{a}{\nu}\udt{T}{\alpha\mu}{\alpha} + \dut{h}{a}{\mu}\udt{T}{\alpha\nu}{\alpha}\,.
\end{equation}
This plays an important role in representing Teleparallel Gravity as a gauge current for a gravitational energy-momentum tensor \cite{deAndrade:2000kr}. \\

\noindent Together, the torsion and superpotential tensors can be combined to produce the so-called torsion scalar
\begin{equation}
T := \dut{S}{a}{\mu\nu}\udt{T}{a}{\mu\nu}\,,
\end{equation}
which is clearly determined through the Weitzenb\"{o}ck connection but can be compared with the Ricci scalar analog that is calculated using the Levi-Civita connection. Coincidentally, it turns out that these scalars are equal up to a total divergence term \cite{Bahamonde:2017ifa,Bahamonde:2015zma}, namely
\begin{equation}\label{TEGR_L}
R=\mathring{R} +T-\frac{2}{h}\partial_{\mu}\left(h\udut{T}{\sigma}{\sigma}{\mu}\right)=0 \quad \Rightarrow \quad \mathring{R} = -T + \frac{2}{h}\partial_{\mu}\left(h\udut{T}{\sigma}{\sigma}{\mu}\right) := -T + B
\end{equation}
where $\mathring{R}$ is the Ricci scalar as determined using the Levi-Civita connection, $R$ is the Ricci scalar as calculated with the Weitzeonb\"{o}ck connection which vanishes, and $h$ is the determinant of the tetrad field, $h = \det\left(\udt{h}{a}{\mu}\right) = \sqrt{-g}$. This fact alone, guarantees that the resulting field equations of a torsion scalar Lagrangian will produce identical field equations as GR while preserving the difference at the level of the Lagrangian and in the theory itself. Secondly, this division means that the second-order contributions of the torsion scalar are not necessarily coupled to the fourth-order terms that result from the boundary scalar $B$. This second point is the source of serious problems in $f(\mathring{R})$ gravity such as ghosts \cite{Capozziello:2018qcp}. \\

\noindent Thus, we can define the so-called TEGR as \cite{Bahamonde:2017wwk}
\begin{equation}
\mathcal{S}_{\rm TEGR} = -\dfrac{1}{2\kappa^2} \int d^4x \: h T + \int d^4x \: h \mathcal{L}_{\rm m}\,,
\end{equation}
where $\kappa^2 = 8\pi G$, and $\mathcal{L}_{\rm m}$ represents the matter Lagrangian. In the Levi-Civita connection paradigm, the field equations of GR are represented by the Einstein tensor through
\begin{equation}
    \mathring{G}_{\mu\nu} := \mathring{R}_{\mu\nu} - \frac{1}{2}g_{\mu\nu}\mathring{R} = \kappa^2 \Theta_{\mu\nu}\,,
\end{equation}
where $\Theta_{\mu\nu}$ is the energy-momentum tensor \cite{misner1973gravitation}. TEGR produces identical field equations, however, these equations appear different since they are described in terms of the tetrad field and spin connection. These are \cite{Krssak:2015oua}
\begin{equation}\label{TEGR_FEs}
    \mathring{G}_{\mu\nu} \equiv \mathcal{G}_{\mu\nu} := h^{-1}h^{a}{}_{\mu}g_{\nu\rho}\partial_\sigma(h S_a{}^{\rho\sigma})-S_{b}{}^{\sigma}{}_{\nu}T^{b}{}_{\sigma\mu}+\frac{1}{4}T g_{\mu\nu}-h^{a}{}_\mu \omega ^{b}{}_{a\sigma}S_{b\nu}{}^{\sigma} = \kappa^2 \Theta_{\mu\nu}\,,
\end{equation}
which is guaranteed to produce the same field equations for any tetrad with the correction-associated spin connection for an equivalent metric tensor ansatz. \\

In a similar manner as the $f(\mathring{R})$ generalization of GR \cite{Faraoni:2008mf,Capozziello:2018qcp}, we can also generalize the gravitational part of this Lagrangian to $f(T,B)$ which forms a larger class of theories than those expressed through $f(\mathring{R})$ gravity (at the level of field equations). This modified Lagrangian has several distinct features such as the subclass of $f(T)$ models producing generally second-order field equations. \\

In general, there are a plethora of potential generalized Teleparallel theories of gravity, as there are in gravity theories based on the Levi-Civita connection. Beyond $f(T,B)$ gravity, some recent progress has been made in Gauss-Bonnet extensions to Teleparallel Gravity \cite{Kofinas:2014owa,Bahamonde:2016kba}. In this case, an extension of TEGR can be written as $-T + f(T,T_G,B_G)$, where $T_G$ represents the Teleparallel equivalent of the Gauss-Bonnet scalar while $B_G$ is the boundary term between the Levi-Civita and Weitzenb\"{o}ck Gauss-Bonnet invariants. Other extensions include considering the trace of the matter Lagrangian \cite{Harko:2014aja,Farrugia:2016pjh}, with other possibilities available.\\

The problem then becomes, how do we build the most general second-order Teleparallel theory of gravity with one scalar field? One way to approach this problem is to consider the irreducible parts with respect to the local Lorentz group through the axial, vector and purely tensorial components, which respectively are~\cite{Bahamonde:2017wwk}
\begin{align}
a_{\mu} &= \frac{1}{6}\epsilon_{\mu\nu\sigma\rho}T^{\nu\sigma\rho}\,, \\
v_{\mu} &= \udt{T}{\sigma}{\sigma\mu}\,,\\
t_{\sigma\mu\nu} &= \frac{1}{2}\left(T_{\sigma\mu\nu}+T_{\mu\sigma\nu}\right) + \frac{1}{6}\left(g_{\nu\sigma}v_{\mu} + g_{\nu\mu}v_{\sigma}\right) - \frac{1}{3}g_{\sigma\mu}v_{\nu}\,,
\end{align}
where $\epsilon_{\mu\nu\sigma\rho}$ is the totally antisymmetric Levi-Civita symbol, and which can then be used to form the scalar invariants
\begin{align}
T_{\text{ax}} &= a_{\mu}a^{\mu} = \frac{1}{18}\left(T_{\sigma\mu\nu}T^{\sigma\mu\nu} - 2T_{\sigma\mu\nu}T^{\mu\sigma\nu}\right)\,,\\
T_{\text{vec}} &=v_\mu v^\mu= \udt{T}{\sigma}{\sigma\mu}\dut{T}{\rho}{\rho\mu}\,, \\
T_{\text{ten}} &= t_{\sigma\mu\nu}t^{\sigma\mu\nu} = \frac{1}{2}\left(T_{\sigma\mu\nu}T^{\sigma\mu\nu} + T_{\sigma\mu\nu}T^{\mu\sigma\nu}\right) - \frac{1}{2}\udt{T}{\sigma}{\sigma\mu}\dut{T}{\rho}{\rho\mu}\,,
\end{align}
that combined as follows form the torsion scalar
\begin{equation}\label{Torsion_sca_scalars}
T = \frac{3}{2} T_{\text{ax}} + \frac{2}{3} T_{\text{ten}} - \frac{2}{3} T_{\text{vec}}\,.
\end{equation}
These three scalars are the most general scalar torsion invariants that are quadratic in the torsion tensor and parity-preserving \cite{PhysRevD.19.3524}, where the parity-violating terms would be
\begin{align}
P_1 &= v^{\mu}a_{\mu}\,,\label{parity_term1}\\
P_2 &= \epsilon_{\mu\nu\sigma\rho}t^{\lambda\mu\nu}\dut{t}{\lambda}{\rho\sigma}\,,\label{parity_term2}
\end{align}
but these are not physical since any Lagrangian scalar should be parity invariant. This means that at second-order, the most general Lagrangian formed by quadratic contractions of torsion that is not parity-violating can be encapsulated in the Lagrangian $f(T_{\text{ax}},T_{\text{vec}},T_{\text{ten}})$. \\


In order to consider a Teleparallel approach to Horndeski gravity, we need a method of covariantizing scalar fields from tangent space to general manifolds. The strong equivalence principle states that for any local Lorentz frame, the spacetime can be described by the Minkowski metric, $\eta_{\mu\nu}$, and acted upon by the partial derivative, $\partial_{\mu}$ \cite{misner1973gravitation}. The Levi-Civita connection, $\mathring{\Gamma}{}^{\sigma}{}_{\mu\nu}$, describes general manifolds that only admit curvature as geometric deformations of neighboring tangent space. In this way, the Levi-Civita connection provides a clear procedure in which to form covariant Lagrangians from their local Lorentz frames, namely through the procedure outlined by
\begin{align}
\eta_{\mu\nu}\,&\rightarrow\,g_{\mu\nu}\,,\nonumber\\
\partial_{\mu}\,&\rightarrow\,\mathring{\nabla}{}_{\mu}\,,\label{GR_equiv_prin}
\end{align}
where the Minkowski metric is raised to the general manifold metric, and the partial derivative is corrected by terms due to parallel transport, which are the well-known Christoffel symbols. \\

Teleparallel gravity is different both physically in that vectors remain parallel at a distance, and in its construction since it is built up from tetrads rather than the metric tensor. To account for this, a different covariantization procedure is needed to raise local Lorentz frame Lagrangians to the general manifold. \\

Stemming from the contorsion tensor relation between the two connections, in Eq.~(\ref{contorsion_def}), it turns out that the coupling prescriptions of both Teleparallel Gravity and GR are equivalent. This results in a covariantization procedure that takes the form (more details in Ref.~\cite{Aldrovandi:2013wha})
\begin{align}
\udt{e}{a}{\mu}\,&\rightarrow\,\udt{h}{a}{\mu},\nonumber\\
\partial_{\mu}\,&\rightarrow\,\mathring{\mathcal{D}}_{\mu}\equiv \mathring{\nabla}{}_{\mu}\,,\label{coup_pres}
\end{align}
where $\mathring{\mathcal{D}}_{\mu}$ is the regular Levi-Civita covariant derivative, $\mathring{\nabla}{}_{\mu}$, calculated using tetrads rather than the metric tensor, but the result is equal. Teleparallel Gravity remains distinct from GR in that it forms gravitational distortions of the general manifold that are entirely described by torsion with vanishing curvature. However, this feature of the theory means that the formulation of the teleparallel analog of Horndeski gravity will, in part, be very natural as compared with its Levi-Civita counterpart.


\section{Horndeski Theory in Teleparallel Gravity}
\label{sec:HorndeskiTeleparallel}

Horndeski's theory of gravity is the most general theory of gravity in a four-dimensional spacetime which is based on contractions of the metric tensor and a single scalar field which leads to second-order field equations in terms of derivatives of the metric~\cite{Horndeski:1974wa}. However, implicit in the derivation of the Lagrangian of Horndeski's theory is its reliance on the Levi-Civita connection~\cite{Kobayashi:2019hrl}. This is not a principle requirement of the approach and can be replaced with other connections or other geometries. \\

By Lovelock's theorem \cite{Lovelock:1971yv}, there is a clear limit to which Lagrangian terms can be used to form a second-order theory. This is extended by contractions with the scalar field in Horndeski's theory. However, this remains finite in the full expansion of the Lagrangian. An example of a Lagrangian that forms a higher-order theory is $f(\mathring{R})$ gravity or Einstein cubic gravity \cite{Bueno:2016xff}.\\

In this section, we discuss the conditions of forming a teleparallel Horndeski analog and the conditions that would necessitate such a formulation.

\subsection{Conditions on a Teleparallel Horndeski Theory\label{conditions}}
In the spirit of Horndeski's original approach, our conditions for forming a teleparallel analog of Horndeski's theory of gravity will be the following \\

\noindent 1. The resulting field equations must be at most second-order in terms of derivatives of the tetrad fields. This is analogous to the condition that the theory is second-order in terms of metric tensor derivatives. The reason for this requirement is to avoid ghost instabilities. \\

\noindent 2. The scalar invariants should not be parity-violating. Using the irreducible parts of the torsion tensor, the full family of contractions will a single scalar field can be considered. However, each scalar must be invariant under parity transformations. \\

\noindent 3. Contractions of the torsion tensor can at most be quadratic. The Lovelock theorem guarantees that no other terms exist in Horndeski's original Lagrangian, but this is not the case in Teleparallel Gravity. Any number of contractions of the irreducible parts of the torsion tensor will result in second-order field equations. This means that an infinite number of terms can be formed in Teleparallel Gravity that give rise to second-order field equations. However, it is unclear how physical such higher-order contributions will be. For this reason, we demand that the contributing scalar invariants of the theory be at most quadratic contractions of the torsion scalar.\\

This is not to say that all of Teleparallel Gravity theories are second-order. Higher-order theories have been formulated and may offer interesting insights such as Ref.~\cite{Otalora:2016dxe}.

\subsection{The most general second-order Lagrangian with one scalar field on a Minkowski background}
To form the most general Lagrangian that adheres to the conditions set out in \S.\ref{conditions}, first the tangent space Lagrangian must be formulated which can then be raised through the coupling prescription to the general manifold. Here, the following conditions on the scalar field are considered (i) the Lagrangian contains at most derivatives second-order in the scalar field; (ii) the Lagrangian is polynomial in second derivatives of the scalar field; (iii) the corresponding field equations are at most second-order in derivatives of the scalar field \cite{Deffayet:2011gz}. Therefore, for a scalar field $\phi$, consider the Lagrangian contributions \cite{Nicolis:2008in}
\begin{align}
L_1 &= \phi\,, \\
L_2 &= X\,, \\
L_3 &= X \bar{\Box}\phi\,, \\
L_4 &= -X\left(\bar{\Box}\phi\right)^2 + \left(\bar{\Box}\phi\right)\partial_{\mu}\phi\partial_{\nu}\phi\partial^{\mu}\partial^{\nu}\phi + X\partial^{\mu}\partial^{\nu}\phi\partial_{\mu}\partial_{\nu}\phi - \partial_{\mu}\phi\partial^{\mu}\partial^{\nu}\phi\partial_{\nu}\partial_{\rho}\phi\partial^{\rho}\phi\,,\\
L_5 &= -2X\left(\bar{\Box}\phi\right)^3-3\left(\bar{\Box}\phi\right)^2\left(\partial_{\mu}\phi\partial_{\nu}\phi\partial^{\mu}\partial^{\nu}\phi\right) + 6X\left(\bar{\Box}\phi\right)\partial_{\mu}\partial_{\nu}\phi\partial^{\mu}\partial^{\nu}\phi\nonumber\\
& + 6\left(\bar{\Box}\phi\right)\partial_{\mu}\phi\partial^{\rho}\phi\partial^{\mu}\partial^{\nu}\phi\partial_{\nu}\partial_{\rho}\phi - 4X\partial^{\nu}\partial_{\mu}\phi\partial^{\rho}\partial_{\nu}\phi\partial^{\mu}\partial_{\rho}\phi\nonumber\\
& + 3\partial_{\mu}\partial_{\nu}\phi\partial^{\mu}\partial^{\nu}\phi\partial_{\rho}\phi\partial_{\lambda}\phi\partial^{\lambda}\partial^{\rho}\phi - 6\partial_{\mu}\phi\partial^{\nu}\partial^{\mu}\phi\partial_{\rho}\partial_{\nu}\phi\partial^{\lambda}\partial^{\rho}\phi\partial_{\lambda}\phi\,,
\end{align}
where $X:=-\frac{1}{2}\partial^{\mu}\phi\partial_{\mu}\phi$ is the kinetic energy, and the full Lagrangian will be
\begin{equation}
L=\sum_{i=1}^5 c_i L_i\,.
\end{equation}
The subscript refers to the number of appearances of the scalar field in each component of the Lagrangian. In this setup, higher appearance Lagrangian terms will appear as total derivatives in four dimensions. Also, the Minkowski d'Alembertian is given by $\bar{\Box}=\partial_{\mu}\partial^{\mu}$ which changes in any gravitational theory since it is a derivative operator. For instance, in GR this takes the form of $\mathring{\nabla}{}_{\mu}\mathring{\nabla}{}^{\mu}$.\\

These components can be compactified by the elimination of total derivative terms through integration by parts, which reduces the Lagrangian to \cite{Kobayashi:2019hrl}
\begin{align}
L = &c_1\phi + c_2 X - c_3 X \bar{\Box}\phi + c_4 X\left[\left(\bar{\Box}\phi\right)^2 - \partial_{\mu}\partial_{\nu}\phi\partial^{\mu}\partial^{\nu}\phi\right] - \nonumber\\
&- c_5 X\left[\left(\bar{\Box}\phi\right)^3 - 3\left(\bar{\Box}\phi\right)\partial_{\mu}\partial_{\nu}\phi\partial^{\mu}\partial^{\nu}\phi + 2\partial_{\mu}\partial_{\nu}\phi\partial^{\nu}\partial^{\lambda}\phi\partial_{\lambda}\partial^{\mu}\phi\right]\,,
\end{align}
where some constants were defined to absorb common factors. In fact, the most general form of this Lagrangian on Minkowski space involves general combinations of the Lagrangian components dependent on the scalar field and the kinetic term~\cite{Deffayet:2011gz}. This is equivalent to raising the linear constant $c_i$ to the arbitrary functions $c_i(\phi,X)$.

\noindent In the Levi-Civita connection approach, the covariantization of the tangent space Lagrangian takes the form \cite{Deffayet:2009wt}
\begin{align}
\mathring{L}{}_2 &= G_2(\phi,X)\,, \label{LC_Horn_Comp1}\\
\mathring{L}{}_3 &= -G_3(\phi,X)\mathring{\Box}\phi\,, \label{LC_Horn_Comp2}\\
\mathring{L}{}_4 &= G_4(\phi,X)\mathring{R} + G_{4,X}(\phi,X)\left[(\mathring{\Box}\phi )^2 - \mathring{\nabla}{}_{\mu}\mathring{\nabla}{}_{\nu}\phi\mathring{\nabla}{}^{\mu}\mathring{\nabla}{}^{\nu}\phi\right]\,, \label{LC_Horn_Comp3}\\
\mathring{L}{}_5 &= G_5(\phi,X)\mathring{G}{}_{\mu\nu}\mathring{\nabla}{}^{\mu}\mathring{\nabla}{}^{\nu}\phi - \frac{1}{6}G_{5,X}(\phi,X)\left[(\mathring{\Box}\phi )^3 + 2\mathring{\nabla}{}_{\nu}\mathring{\nabla}{}_{\mu}\phi\mathring{\nabla}{}^{\nu}\mathring{\nabla}{}^{\lambda}\phi\mathring{\nabla}{}_{\lambda}\mathring{\nabla}{}^{\mu}\phi - 3\mathring{\Box}\phi \mathring{\nabla}{}_{\mu}\mathring{\nabla}{}_{\nu}\phi\mathring{\nabla}{}^{\mu}\mathring{\nabla}{}^{\nu}\phi\right]\,, \label{LC_Horn_Comp4}
\end{align}
where $G_i$ are arbitrary functions of the scalar field and the kinetic term, the first and second Lagrangian parts have been combined to form the function $G_2(\phi,X)$, and kinetic term derivatives are represented by $G_{i,X}=\partial G_i/\partial X$. The kinetic term is calculated using the regular Levi-Civita connection covariant derivative, $X=-\frac{1}{2}\mathring{\nabla}{}_{\mu}\phi\mathring{\nabla}{}^{\mu}\phi$. The d'Alembertian operator takes the usual form $\mathring{\Box}:=\mathring{\nabla}{}_{\mu}\mathring{\nabla}{}^{\mu}$, where the explicit dependence on the Levi-Civita connection is shown by the overcircle as specified in Eq.~(\ref{TEGR_L}). This notation has also been extended to the derivative operators. Also, the kinetic term is determined through covariant derivatives with respect to the Levi-Civita connection, which is retained throughout the rest of the work (which is identical to teleparallel kinetic term $X=-\frac{1}{2}\mathring{\mathcal{D}}{}_{\mu}\phi\mathring{\mathcal{D}}{}^{\mu}\phi$). The reasoning for this retention is the teleparallel coupling prescription, as is done in other works \cite{Abedi:2018lkr}.\\

It should also be noted that the $\mathring{L}_4$ and $\mathring{L}_5$ Lagrangian parts contain not only the Levi-Civita covariantization terms but also correction terms. The Levi-Civita connection covariantization procedure leads to higher-order derivative terms which must be corrected for through curvature correction terms which is how these additional Lagrangian terms arise \cite{Deffayet:2009wt,Kobayashi:2019hrl}. In this form, Horndeski gravity is strongly limited by recent gravitational wave observations through the speed of propagation constraints \cite{Ezquiaga:2017ekz} which has set stringent limits to the form of the $\mathring{L}_4$ and $\mathring{L}_5$ components \cite{Sakstein:2017xjx,Creminelli:2017sry}. However, several new avenues are being advanced to circumvent this problem within the Levi-Civita connection paradigm \cite{Gleyzes:2014dya,Kobayashi:2019hrl} which have had interesting results. \\

In this work, we aim to keep to the original spirit of Horndeski's approach while changing the mode in which gravity is characterized through the Weitzenb\"{o}ck connection. To achieve this goal, we must consider the covariantization procedure through the coupling procedure described in \S.\ref{sec:Intro_Tele_Grav}, while adhering to the teleparallel analog of the conditions of the theory in \S.\ref{conditions}.

\subsection{The Teleparallel Horndeski Theory}
To form the Teleparallel Gravity analog of the Levi-Civita form of Horndeski's theory, we must use the coupling procedure on each of the Lagrangian terms in the Minkowski background space Lagrangian. Given that Minkowski space has no gravitational effects, this starting point remains invariant in both formulations of gravity. In the Levi-Civita setup, Horndeski gravity leads to the Lagrangian components in Eqs.~(\ref{LC_Horn_Comp1})-(\ref{LC_Horn_Comp4}). Similarly, in Teleparallel Gravity, the same number of contributions emerge, however these will differ to varying degrees due to the change in connection which will retain all the standard Horndeski terms due to the coupling prescription nature, and actually add further terms due to the nature of Lovelock's theorem in Teleparallel Gravity. We consider each contribution in turn and describe how they change in this setting. \\

Firstly, as in the standard setup, the first two Minkowski space Lagrangian components are combined to form an arbitrary function of both the scalar field, $\phi$ and kinetic term, $X$. This gives rise to the usual Lagrangian component
\begin{equation}
    \mathcal{L}_2 = G_2(\phi,X)\,,\label{L2}
\end{equation}
where $X$ continues to be dependent on the Levi-Civita connection but is calculated through the tetrads using the $\mathring{\mathcal{D}}$ operator, i.e. $X=-\frac{1}{2}\mathring{\mathcal{D}}{}^{\mu}\phi\mathring{\mathcal{D}}{}_{\mu}\phi=-\frac{1}{2}\partial_{\mu}\phi\partial^{\mu}\phi$. Similarly, the coupling prescription for Teleparallel Gravity gives the same form of the Lagrangian component for $\mathcal{L}_3$, which turns out to be
\begin{equation}
    \mathcal{L}_3 = G_3(\phi,X)\Box\phi\,,\label{L3}
\end{equation}
where the d'Alembertian operator retains its Levi-Civita connection form due to the coupling prescription, but is determined using the tetrad fields, i.e. $\Box:=\mathring{\mathcal{D}}{}_{\mu}\mathring{\mathcal{D}}{}^{\mu}=\mathring{\nabla}{}_{\mu}\mathring{\nabla}{}^{\mu}$. The $\mathcal{L}_2$ and $\mathcal{L}_3$ cannot produce field equations higher than second-order in terms of derivatives of the tetrad or scalar field because the first term has no derivatives at the Lagrangian level and the second forms a Dvali–Gabadadze–Porrati (DGP) term \cite{Luty:2003vm}. \\

On the same line of reasoning, the $\mathring{L}_4$ standard component in Eq.~(\ref{LC_Horn_Comp3}) features a correction term from which the Ricci scalar enters the theory. Stemming from the coupling prescription, the teleparallel scenario will also require a correction term to preserve the second-order nature of the resulting field equations. In Ref.~\cite{Deffayet:2009wt}, the higher-order contributions to the field equations are negated by considering several Lagrangian terms with containing contractions of the Riemann tensor with scalar field derivatives. Using integration by parts, this results in the single term correction of the Ricci scalar. Following the equivalent procedure, it follows that the same standard Horndeski component will turn out to be
\begin{equation}\label{L4tele}
\mathcal{L}_4 = G_4(\phi,X)\left(-T+B\right) + G_{4,X}(\phi,X)\left[\left(\Box\phi\right)^2 - \phi_{;\mu\nu}\phi^{;\mu\nu}\right]\,,
\end{equation}
where semicolon denotes the Levi-Civita covariant derivative calculated with the tetrad field, i.e. $\mathring{\mathcal{D}}{}_{\mu}$. In Teleparallel Gravity the torsion and Ricci scalars are equal up to a total derivative term, as shown in Eq.~(\ref{TEGR_L}), which is the physical reasoning behind the apparent symmetry between the standard Horndeski terms and their Teleparallel Gravity analog. In what follows, we will introduce a new Lagrangian component that is only apparent in the Teleparallel Gravity paradigm since more scalar invariants can be constructed in this theory that produce second-order field equations. For this reason, the nonminimally coupled torsion scalar will also appear as an unconstrained contribution coupled with at most first-order derivatives of the scalar field. \\

The final term to contribute in four dimensions is $L_5$, which in the standard Horndeski theory gives a covariantized contribution shown in Eq.~(\ref{LC_Horn_Comp4}). In the Levi-Civita connection framework, the Minkowski contribution gives rise to higher-order derivatives in the metric which are mended by adding the Einstein tensor term to eliminate such contributions. For the teleparallel case, as for the $\mathcal{L}_4$ correction, the combination of the coupling prescription and the equivalence of the torsion and Ricci scalars results trivially in the same correction term after an integration by parts procedure. However, the Einstein tensor must be replaced with its teleparallel equivalent through the tensor $\mathcal{G}_{\mu\nu}$ in Eq.~(\ref{TEGR_FEs}). This results in the Lagrangian component
\begin{equation}
    \mathcal{L}_5 = G_5(\phi,X)\mathcal{G}_{\mu\nu}\phi^{;\mu\nu} - \frac{1}{6}G_{5,X}(\phi,X)\left[\left(\Box\phi\right)^3 + 2\dut{\phi}{;\mu}{\nu}\dut{\phi}{;\nu}{\alpha}\dut{\phi}{;\alpha}{\mu} - 3\phi_{;\mu\nu}\phi^{\mu\nu}\left(\Box\phi\right)\right]\,,\label{L5}
\end{equation}
which completes the correspondence of each of the standard Horndeski Lagrangian terms. However, Teleparallel Gravity offers more scalar invariants than GR since Lovelock's theorem is weakened under the Weitzenb\"{o}ck connection. This leads to a further Lagrangian contribution, $\mathcal{L}_{\text{Tele}}$, which will contain the full family of further contributions that preserve the second-order field equations. \\

As discussed in \S.\ref{sec:Intro_Tele_Grav}, taking quadratic contractions of the torsion tensor, the most general Lagrangian of Teleparallel Gravity turns out to be $f(T_{\text{ax}},T_{\text{vec}},T_{\text{ten}})$ neglecting the unphysical parity-violating contributions in Eqs.~(\ref{parity_term1})-(\ref{parity_term2}). This means that conforming to the third condition in \S.\ref{conditions} means that these scalar invariants must also be included in the $\mathcal{L}_2$ of Teleparallel Gravity. \\

Teleparallel Gravity is distinctly different to GR in that it remains second-order in derivative operators on the tetrad field for arbitrary functions of the scalar contributions to the TEGR Lagrangian through $f(T_{\text{ax}},T_{\text{vec}},T_{\text{ten}})$. This creates a second set of scalars that must be included in the $\mathcal{L}_{\text{Tele}}$ component, namely the irreducible parts of the torsion tensor contracted with covariant derivatives of the scalar field. Again, given the form of the coupling procedure, this will mean Levi-Civita derivatives of the scalar field, $\mathring{\mathcal{D}}{}_{\mu}\phi=\mathring{\nabla}{}_{\mu}\phi=\partial_{\mu}\phi$. Given the definition of the torsion scalar in Eq.~(\ref{Torsion_sca_scalars}), we can equivalently describe this family of functions as $f(T_{\text{ax}},T_{\text{vec}},T)$ where the correspondence with the literature is more apparent.\\

\noindent Thus, the full set of linear contractions of the torsion tensor irreducibles can be encapsulated in the scalars
\begin{align}
I_1 &= t^{\mu\nu\sigma}\phi_{;\mu}\phi_{;\nu}\phi_{;\sigma}\,,\\
I_2 &= v^{\mu}\phi_{;\mu}\,,\label{S2}\\
I_3 &= a^{\mu}\phi_{;\mu}\,.
\end{align}
Given the symmetry of the purely torsional part in the first two indices, $t_{\mu\nu\sigma}=t_{\nu\mu\sigma}$, and the fact that any pair of contractions of this part vanishes, $\udt{t}{\sigma\mu}{\sigma}=0=\udut{t}{\sigma}{\sigma}{\mu}=\udt{t}{\mu\sigma}{\sigma}$, renders these the full set scalars formed by linear contractions of the torsion tensor. However, due to the anti-symmetry of the torsion tensor in its last two indices, it can easily be shown that $I_1$ vanishes. Also, since we require parity-preserving scalars, $I_3$ cannot be considered because it does not feature this property. More generally, scalars made of odd appearances of the axial irreducible part of the torsion tensor form parity-violating scalars, which will be important for the full set of quadratic contractions of the torsion tensor irreducibles with derivatives of the scalar field. While second-order at Lagrangian level, the set of permutations of contractions of the purely tensorial part with second-order derivatives of the scalar field result in higher-order derivatives in the resulting field equations. \\

In line with the third condition of \S.\ref{conditions}, we now consider the complete set of quadratic contractions of the torsion tensor that involve first derivatives of the scalar field. Keeping only those scalars that are invariant under parity transformations, the resulting contributions are
\begin{align}
    J_1 &= a^{\mu}a^{\nu}\phi_{;\mu}\phi_{;\nu}\,, \\
    J_2 &= v^{\mu}v^{\nu}\phi_{;\mu}\phi_{;\nu}\,, \\
    J_3 &= v_{\sigma}t^{\sigma\mu\nu}\phi_{;\mu}\phi_{;\nu}\,, \\
    J_4 &= v_{\mu}t^{\sigma\mu\nu}\phi_{;\sigma}\phi_{;\nu}\,, \\
    J_5 &= t^{\sigma\mu\nu}\dudt{t}{\sigma}{\bar{\mu}}{\nu}\phi_{;\mu}\phi_{;\bar{\mu}}\,, \\
    J_6 &= t^{\sigma\mu\nu}\dut{t}{\sigma}{\bar{\mu}\bar{\nu}}\phi_{;\mu}\phi_{;\nu}\phi_{;\bar{\mu}}\phi_{;\bar{\nu}}\,, \\
    J_7 &= t^{\sigma\mu\nu}\udt{t}{\bar{\sigma}\bar{\mu}}{\sigma}\phi_{;\mu}\phi_{;\nu}\phi_{;\bar{\sigma}}\phi_{;\bar{\mu}}\,, \\
    J_8 &= t^{\sigma\mu\nu}\dut{t}{\sigma\mu}{\bar{\nu}}\phi_{;\nu}\phi_{;\bar{\nu}}\,, \\
    J_9 &= t^{\sigma\mu\nu}t^{\bar{\sigma}\bar{\mu}\bar{\nu}}\phi_{;\sigma}\phi_{;\mu}\phi_{;\nu}\phi_{;\bar{\sigma}}\phi_{;\bar{\mu}}\phi_{;\bar{\nu}}\,, \\
    J_{10}&=\epsilon^{\mu}{}_{\nu\rho\sigma}a^\nu t^{\alpha\rho\sigma}\phi_{;\mu}\phi_{;\alpha}\,.
\end{align}
The set of contractions involving the full set of permutations with second derivatives of the scalar field produce higher-order scalar invariants and are thus not included. Also, as with $I_1$, the anti-symmetry of the last two indices of the torsion tensor means that $J_9$ vanishes. One can also notice that $J_2=I_2^2$, $J_3=J_4$ and $J_7=-2J_6$ due to the symmetric property $t_{\lambda\mu\nu}+t_{\mu\nu\lambda}+t_{\nu\lambda\mu}=0$. Then, there are only seven extra independent scalars containing scalar field derivatives and torsion up to quadratic contractions of torsion. Therefore, Teleparallel Gravity produces the further Lagrangian contribution
\begin{equation}\label{L2tele}
    \mathcal{L}_{\text{Tele}}=G_{\text{Tele}}(\phi,X,T,T_{\text{ax}},T_{\text{vec}},I_2,J_1,J_3,J_5,J_6,J_8,J_{10}),
\end{equation}
which forms the full set of scalar invariants that produce second-order field equations beyond the standard Horndeski components. Notice that the boundary term, $B$, does not contribute. While $B$ is second-order in the Lagrangian, it produces the fourth-order elements of the theory. This is the source of all the fourth-order terms in $f(\mathring{R})$ gravity. For this reason, it is not included in this Lagrangian component. \\

The extra Lagrangian term presented in Eq.~\eqref{L2tele} is very interesting because it naturally contains $f(T)$ gravity as a subclass of the Teleparallel Gravity analog of Horndeski theory which is different from the standard scenario where $f(\mathring{R})$ gravity produces fourth-order contributions for nontrivial modifications of the Lagrangian. $f(T)$ gravity is completely second-order and satisfies the conditions to produce a Horndeski analog within the teleparallel context, which means that this new larger class of theories contains within it several positive features such as the identical polarization modes of GR~\cite{Farrugia:2018gyz}, and the resolution of the cosmological $H_0$ tension~\cite{Nunes:2018xbm}. Last but not least, the new theory should also be viable in the weak field limit. The (post-)Newtonian limit of the standard Horndeski is known~\cite{Hohmann:2015kra} and is determined by 15 constant parameters, that can be constrained depending on whether the scalar field is massive or not. The contribution of the new Eq.~(\ref{L2tele}) in the Teleparallel analog will be studied in Ref.~\cite{PPNinprep}.\\

\noindent Hence, the Teleparallel analog of Horndeski's theory of gravity is given by the Lagrangian
\begin{equation}\label{Lagrangian}
    \mathcal{L}=\sum_{i=2}^5 \mathcal{L}_{i} + \mathcal{L}_{\text{Tele}}\,,
\end{equation}
which is the most general theory with one scalar field leading to second-order field equations in terms of derivatives with respect to the tetrad or scalar field, and containing scalar invariants that are at most quadratic in torsion tensor contractions. Here, the Lagrangians $\mathcal{L}_i$, for $i=2,..,5$, are given by Eqs.~\eqref{L2}-\eqref{L5}, respectively. The Lagrangian in Eq.~(\ref{Lagrangian}) is the Teleparallel Gravity analog of the standard Horndeski theory given in Eqs.~(\ref{LC_Horn_Comp1})-(\ref{LC_Horn_Comp4}), but to recover the exact form the new component, $\mathcal{L}_{\text{Tele}}$ would need to vanish. Despite this fact, one should also notice that an overlap exists where the identical potential coupling between the torsion scalar can be produced in $\mathcal{L}_4$ and $\mathcal{L}_{\text{Tele}}$. This can be eliminated by redefining the new Horndeski Lagrangian term as $\tilde{G}_{\rm Tele}=G_{\rm Tele}+TG_4(\phi,X)$, so that scalar field couplings with the boundary term alone would then be possible in Eq.~\eqref{L4tele}. In fact, studies of the coupling of the boundary term with the scalar field already exist in the literature \cite{Bahamonde:2019gjk,Bahamonde:2015hza,Zubair:2016uhx}.

\section{Cosmology in Teleparallel Horndeski Theory}
\label{sec:CosmoTeleDeski}

In this section, we will study flat FLRW cosmology for teleparallel Horndeski theory. In this case, the metric in Cartesian coordinates is
\begin{equation}
ds^2=-N(t)^2dt^2+a(t)^2(dx^2+dy^2+dz^2)\,,
\end{equation}
where $N$ is the lapse function and $a(t)$ is the scale factor. Without losing generality, one can take a zero spin connection gauge, $\omega^{a}{}_{b\mu}=0$, and write down the following diagonal tetrad~\cite{Krssak:2018ywd}
\begin{equation}
h^a{}_{\mu}=\textrm{diag}(N(t),a(t),a(t),a(t))\,.
\end{equation}
For this spacetime, only the vectorial part of the torsion tensor is nonzero, which is explicitly given by
\begin{equation}
v_{\mu}=(-3H,0,0,0)\,,
\end{equation}
where $H=\dot{a}/a$ is the Hubble parameter. Thus, when one is considering contractions of the torsion tensor without considering the scalar field, only  $T_{\rm vec}$ is nonzero and reads
\begin{equation}
T_{\rm vec}=\frac{-9H^2}{N^2}\,.
\end{equation}
Clearly, the torsion scalar only depends on this quantity (see Eq.~\eqref{Torsion_sca_scalars}) giving $T=(-2/3)T_{\rm vec}=6H^2/N^2$. There are no other possible nonzero contractions of the torsion tensor than the above scalar in flat FLRW cosmology. Thus, it is equivalent to take $T$ in a function in a Lagrangian instead of having $T_{\rm vec}$ since they are only related by a constant. Since mainly all the literature in teleparallel have worked on theories constructed by the torsion scalar $T$, it is convenient to choose this quantity in the Lagrangian.

When one is considering couplings between the torsion tensor and $\phi$, then in flat FLRW, only the couplings related to the vectorial part of the torsion tensor $v^\mu$ and derivatives of $\phi$ will be nonzero. If one only considers up to quadratic contractions of torsion (See condition 3 in~\S.\ref{conditions}), the only scalar that is nonzero is then $I_2=v^\mu \phi_{;\mu}$. If ones relaxes the condition of only considering terms constructed up to quadratic contractions of torsion, one can also incorporate the following terms
\begin{equation}
I_{2,n}=\underset{(n \ \textrm{times})}{v^{\mu}v^{\nu}v^{\alpha}\cdots v^{\epsilon}}\underset{(n \ \textrm{times})}{(\mathring{\mathcal{D}}{}_{\mu}\phi)(\mathring{\mathcal{D}}{}_{\nu}\phi) (\mathring{\mathcal{D}}{}_{\alpha}) \cdots (\mathring{\mathcal{D}}{}_{\epsilon}\phi) }\,,
\end{equation}
where $n \in \mathcal{N}$. It turns out that for flat FLRW cosmology, for any contraction $n$ it is possible to get that those scalars behave as
\begin{equation}
I_{2,n}=\Big(\frac{3H\dot{\phi}}{N^2}\Big)^n=I_2^n\,.
\end{equation}
Since these scalars only depend on $I_2$ with different exponents $n$, it is sufficient to add $I_2$ as an argument of a function in a Lagrangian to get all the possible terms that one can construct from them. Therefore, in flat FLRW cosmology, it is possible to write down an action with finite scalars even considering higher-order contractions of the torsion tensor. Again, this happens only because the axial and tensorial parts of torsion are zero for this case.

\noindent Therefore, for flat FLRW, the most general Lagrangian $\mathcal{L}_{\rm Tele}$ that can be constructed by the torsion tensor and one scalar field which leads second-order field equations can be written as
\begin{equation}
\mathcal{L}_{\rm Tele}=\tilde{G}_2(\phi,X,T,I_{2})\,.
\end{equation}
Since the modified FLRW equations are long, we will split each contribution for each Lagrangian piece $\mathcal{L}_{i}$. Clearly, only $\mathcal{L}_{\rm Tele}$ differs from the standard Horndeski Lagrangian constructed using General Relativity and curvature. By doing variations with respect to the lapse function $N(t)$ one gets the first Friedmann equation which can be split as
\begin{equation}
{\cal E}_{\rm Tele}+\sum_{i=2}^5 {\cal E}_i =0\,,
\end{equation}
where
\begin{eqnarray}
{\cal E}_{\rm Tele}&=&6 H\dot{\phi}\tilde{G}_{2,I_2}+12 H^2 \tilde{G}_{2,T}+2X \tilde{G}_{2,X}-\tilde{G}_{2}\,,\\
{\cal E}_2&=&2XG_{2,X}-G_2\,,\label{F1b}\\
{\cal E}_3&=&6X\dot\phi HG_{3,X}-2XG_{3,\phi}\,,
\\
{\cal E}_4&=&-6H^2G_4+24H^2X(G_{4,X}+XG_{4,XX})
-12HX\dot\phi G_{4,\phi X}-6H\dot\phi G_{4,\phi }\,, \label{F1c}
\\
{\cal E}_5&=&2H^3X\dot\phi\left(5G_{5,X}+2XG_{5,XX}\right)\label{F1d}
-6H^2X\left(3G_{5,\phi}+2XG_{5,\phi X}\right)\,,
\end{eqnarray}
where $G_{2,X}=\partial G_2/\partial X$, $G_{5,XX}=\partial^2 G_5/\partial X^2$ and so on, therefore commas denote differentiation. Each subscript represents the contribution of each Lagrangian \eqref{Lagrangian} to the first FLRW equation. Eqs.~\eqref{F1b}-\eqref{F1d} are identical to Eqs.~(3.2)--(3.5) reported in \cite{Kobayashi:2011nu}, however now there is another contribution from ${\cal E}_{\rm Tele}$ due to teleparallel Horndeski cosmology.

\noindent Now, if one varies the action with respect to the scale factor $a(t)$ one gets the following set of equations,
\begin{eqnarray}
{\cal P}_{\rm Tele}+\sum_{i=2}^5{\cal P}_i=0\,,  \label{P}
\end{eqnarray}
where
\begin{eqnarray}
{\cal P}_{\rm Tele}&=&-3 H\dot{\phi}\tilde{G}_{2,I_2}-12 H^2\tilde{G}_{2,T}-\frac{d}{dt}\Big(4H \tilde{G}_{2,T}+\dot{\phi}\,\tilde{G}_{2,I_2}\Big)+\tilde{G}_2\,,
\\
{\cal P}_2&=&G_2\,,\\
{\cal P}_3&=&-2X\left(G_{3,\phi}+\ddot\phi G_{3,X} \right) \,,
\\
{\cal P}_4&=&2\left(3H^2+2\dot H\right) G_4
-12 H^2 XG_{4,X}-4H\dot X G_{4,X}
-8\dot HXG_{4,X}-8HX\dot X G_{4,XX}
\nonumber\\&&
+2\left(\ddot\phi+2H\dot\phi\right) G_{4,\phi}
+4XG_{4,\phi\phi}
+4X\left(\ddot\phi-2H\dot\phi\right) G_{4,\phi X}\,,
\\
{\cal P}_5&=&-2X\left(2H^3\dot\phi+2H\dot H\dot\phi+3H^2\ddot\phi\right)G_{5,X}
-4H^2X^2\ddot\phi G_{5,XX}
\nonumber\\&&
+4HX\left(\dot X-HX\right)G_{5,\phi X}
+2\left[2\frac{d}{dt}\left(HX\right)+3H^2X\right]G_{5,\phi}
+4HX\dot\phi G_{5,\phi\phi}\,.
\end{eqnarray}
Finally, by taking variations with respect to the scalar field one gets the following modified Klein Gordon equation
\begin{eqnarray}
\frac{1}{a^3}\frac{d}{dt}\Big[a^3 (J+J_{\rm Tele})\Big]=P_{\phi}+P_{\rm Tele},
\end{eqnarray}
where $J$ and $P_{\phi}$ are the standard terms in the modified Klein Gordon equation in standard Horndeski theory that comes from the Lagrangians $\mathcal{L}_i$, where $i=2,..,5$, namely~\cite{Kobayashi:2011nu}
\begin{eqnarray}
J&=&\dot\phi G_{2,X}+6HXG_{3,X}-2\dot\phi G_{3,\phi}
+6H^2\dot\phi\left(G_{4,X}+2XG_{4,XX}\right)-12HXG_{4,\phi X}
\nonumber\\&&
+2H^3X\left(3G_{5,X}+2XG_{5,XX}\right)
-6H^2\dot\phi\left(G_{5,\phi}+XG_{5,\phi X}\right)\,,\\
P_{\phi}&=&G_{2,\phi}-2X\left(G_{3,\phi\phi}+\ddot\phi G_{3,\phi X}\right)
+6\left(2H^2+\dot H\right)G_{4,\phi}
+6H\left(\dot X+2HX\right)G_{4,\phi X}
\nonumber\\&&
-6H^2XG_{5,\phi\phi}+2H^3X\dot\phi G_{5,\phi X}\,,
\end{eqnarray}
and $J_{\rm Tele}$ and $P_{\rm Tele}$ are new terms related to the teleparallel Horndeski, given by
\begin{eqnarray}
J_{\rm Tele}&=&\dot{\phi}\tilde{G}_{2,X}\,,\\
P_{\rm Tele}&=&-9 H^2\tilde{G}_{2,I_2}+\tilde{G}_{2,\phi}-3  \frac{d}{dt}\left(H\tilde{G}_{2,I_2}\right)\,.
\end{eqnarray}
It is important to emphasize again that teleparallel Horndeski cosmology also contains the standard Horndeski cosmology since if one takes $\tilde{G}_2=0$, one recovers the latter theory. Thus, one can conclude that a teleparallel Horndeski version is richer than the standard Horndeski one. \\

Moreover, since $f(T)$ is a subclass of teleparallel Horndeski, one can also conclude that teleparallel Horndeski can explain both dark energy and inflation~\cite{Nunes:2018evm,Cai:2015emx,Dent:2011zz,Bamba:2010wb,Gonzalez-Espinoza:2019ajd}, can have bounces cosmological solutions~\cite{Bamba:2016gbu,Cai:2011tc} and also can alleviate the $H_0$ tension~\cite{Nunes:2018xbm}. Since this theory also contains the Teleparallel scalar-tensor theories studied in~\cite{Bahamonde:2015hza,Geng:2011aj}, it can describe a crossing of the phantom barrier, quintessencelike or phantomlike behavior, and also can describe a late time accelerating attractor solution without requiring any fine-tuning of the parameters. Then, teleparallel Horndeski has an enlarged number of theories as compared with standard Horndeski that can explain the cosmological observations without introducing a cosmological constant. It is then important to analyze how these theories can pass the constraints coming from gravitational waves to become more reliable theories. Thus, since teleparallel Horndeski contains many important subclasses theories such as the ones related to the Teleparallel part and also the ones coming from its standard Horndeski part, it is a potential cosmological viable model that is important to study in full detail. Since this is not the real aim of this work, this study will be carried out in the future.

\section{Conclusions}
\label{sec:Conclusions}

In this work, we have introduced the analog of Horndeski gravity in the teleparallel framework. It is well known that a large part of Horndeski's theory in the curvature case has been eliminated from the GW170817 event. That is why we formulated its analog considering torsion as the mediator for gravity. After setting a specific set of conditions, based on which we built our theory, we showed that the teleparallel Horndeski theory presents a  much richer phenomenology than the standard Horndeski. Specifically, because of the form of the torsion tensor and its irreducible decomposition, one can construct a full set of 14 scalar invariants that appear in this framework. Based on that, we see that terms of the $\mathcal{L}_4$, i.e. Eq.~\eqref{L4tele}, that in the standard Horndeski theory are severely constrained, could in this case survive through the Lagrangian contribution in Eq.~\eqref{L2tele}. It should be emphasize here that the standard Horndeski gravity, constructed from curvature with the Levi-Civita connection, is a subcase of the teleparallel Horndeski gravity. One recovers the standard case by setting $\mathcal{L}_{\rm Tele}=0$, so that, teleparallel Horndeski has a richer structure than the standard Horndeski gravity theory. 

Moreover, $f(T)$ gravity which has been widely studied in the literature, is also a subcase of teleparallel Horndeski gravity. In standard Horndeski, $f(\mathring{R})$ does not appear in the Lagrangian since this theory give rise to fourth-order derivatives in the field equations. This is remarkable since it seems that teleparallel Horndeski exhibits a more natural scalar field extension to consider than its standard form based from curvature. 

To fully depict the impact of a Teleparallel Gravity analog of the standard Horndeski theory of gravity, we show its relation to other modified teleparallel theories in Fig.~(\ref{fig}). Here, we show how each subclass of theories produces different avenues for extended Teleparallel Gravity, while also showing the relation to the standard Horndeski gravity theory, and some of its subclasses. On each case, we have also included some important references related to those theories. The top part of the figure ($\mathcal{L}_{\rm Tele}=0$), labelled with clouds, represents theories that are related to the standard Horndeski theory and, hence, they can be written down only with quantities related to the curvature computed with the Levi-Civita connection. Even though this quantity does not appear explicitly in the Teleparallel Lagrangian, due to Eqs.~\eqref{TEGR_L} and \eqref{TEGR_FEs}, one notices that they can be rewritten only with quantities computed with the Levi-Civita connection. In those cases, the spin connection $\omega^{a}{}_{b\mu}$ disappears in the field equations as it needs to be for theories constructed by modifying General Relativity. Moreover, in those theories, one only needs the metric (and not the tetrads) to fully determine the important geometrical quantities. This is of course fully consistent with the standard curvature-based theories of gravity constructed from GR. At the bottom part of the figure ($\mathcal{L}_{\rm Tele}\neq 0$), represented with blocks, we have depicted some Teleparallel theories that have been studied in the past and how one can recover them by taking the specific limits for the Lagrangian. It should be noted that we have not depicted all the possible theories that one could construct from teleparallel Horndeski. Our aim was to show how the most important theories studied in the literature are related and how to obtain them by assuming some ansatz for the Lagrangian. Further, it is possible to construct a large number of new Teleparallel theories due to the inclusion of the Lagrangian $\mathcal{L}_{\rm Tele}$. It would be interesting to analyze those new theories in full detail both from the cosmological and astrophysical point of view. To emphasize the fact the TEGR and GR have the same field equations, we have also included a small box labeling this. It is important to remark that for theories with $\mathcal{L}_{\rm Tele}\neq 0$, one cannot fully determine all the Lagrangian quantities with only the metric since the terms appearing in that Lagrangian are constructed from Teleparallel Gravity, so that, tetrads and spin connection must be taken into account. Another important point to remark is that teleparallel Horndeski is covariant under LLTs. To understand this clearly, see the review~\cite{Krssak:2018ywd}, where it is explained how to model modified Teleparallel theories of gravity without loosing the Lorentz invariance.

All in all, we think that this paper will significantly contribute in the modified gravity community because it introduces a new general theory, that is the teleparallel Horndeski gravity. In an upcoming work we plan to study the effect of the gravitational wave event (and its electromagnetic counterpart) in this theory, in order to specifically see if indeed some of the excluded terms of the standard Horndeski will survive in this framework.

\newpage
\begin{figure}[H]\label{Tele_Map}
	\centering
\small{\begin{tikzpicture}
[auto,
decision/.style={diamond, draw=blue, thick, fill=blue!20,
	text width=8em,align=flush center,
	inner sep=1pt},
block/.style ={rectangle, draw=blue, thick, fill=blue!20,
	text width=8em,align=center, rounded corners,
	minimum height=3em},
line/.style ={draw, thick, -latex',shorten >=2pt},
cloud/.style ={draw=red, thick, ellipse,text width=8em,fill=red!20,	text width=6em,align=center,	minimum height=3em}]
\matrix [column sep=12mm,row sep=30mm]
{&	&\node [cloud] (GR) {GR};\\[-11ex]
	 &
	\node [cloud] (expert4) {Kinetic braiding~\cite{Deffayet:2010qz}}; &
	\node [cloud] (expert) {Generalized Brans-Dicke~\cite{Brans:1961sx,Perrotta:1999am}}; &
	\node [cloud] (init) {k-essence, quintessence~\cite{Copeland:2006wr}};  \\[-8ex]
	\node [cloud] (expert2) {Quartic couplings};&	& \node [cloud] (identify) {Horndeski\newline~\cite{Horndeski:1974wa,Kobayashi:2011nu}}; & \\[-15ex]
		\node [cloud] (GR2) {GR};
		&\node [block] (TEGR2) {TEGR};
		& \node [decision] (evaluate) {Teleparallel Horndeski};\\[-21ex]
	\node [block] (decide8) {Cubic Teleparallel \newline\cite{Gonzalez-Espinoza:2019ajd}};	&	& \node  (evaluate2) {}; & \node [block] (decide5) {Conformal Teleparallel~\cite{Maluf:2011kf}};\\[6ex]
	\node [block] (decide3) {Kinetic Teleparallel~\cite{Hohmann:2018dqh,Abedi:2015cya}};	& \node [block] (decide) {Non-minimally couplings between $\phi$ and $T$ and $B$~\cite{Zubair:2016uhx}}; &\node [block] (decide4) {$f(T,T_{\rm ax},T_{\rm vec})$\cite{Bahamonde:2017wwk}}; & \\[-4ex]
	\node [block] (stop2) {Couplings with $T$~\cite{Geng:2011aj}};	& \node [block] (stop) {Couplings with $B$~\cite{Bahamonde:2015hza}}; & \node [block] (decide6) {$f(T)$~\cite{Ferraro:2006jd,Cai:2015emx}}; & \node [block] (decide7) {New General Relativity~\cite{PhysRevD.19.3524}};\\[-11ex]
&	&\node [block] (TEGR) {TEGR};\\
};
\begin{scope}[-stealth,every path/.style]
\path (identify) -- (init);
\path (GR2) edge node [right] {}  (TEGR2);
\path (TEGR2) edge node [right] {}  (GR2);
\draw [rounded corners=0.8cm] (-8.5,0.5) rectangle ++(8,2.5) node [midway] {};
\node[anchor=west,text width=5cm] (note1) at (-7.5,1.1) {
   GR and TEGR are equivalent at the level of their field equations};
\path (evaluate) edge node [right] {$\mathcal{L}_{\rm Tele}=0$}  (identify);
\path (evaluate) edge node [right] {$\mathcal{L}_{\rm Tele}\neq0$} (evaluate2);
\path (evaluate2) edge node [above,sloped] {\scriptsize{$\hspace{-1cm}\tilde{G}_{\rm Tele}=(\tilde{G}_4(\phi)+F(\phi))T$}} (decide);
\path (evaluate2) edge node [below,sloped] {\scriptsize{$G_5=G_3=0,G_4=\tilde{G}_4(\phi)$}} (decide);
\path (evaluate2) edge node [above] {$G_4=G_5=0, G_{\rm Tele}=F(\phi)T$ } (decide8);
	\path (evaluate2) edge node [sloped,above] {$G_3=G_4=G_5=0$ and $G_{\rm Tele}=F(\phi,X)T$} (decide3);
	\path (evaluate2) edge node [above,sloped] {$G_2=G_3=G_4=G_5=0$} (decide4);
	
	\path (decide8) edge[bend right=40] node [above,sloped] {$G_3=0$} (stop2);
	
		\path (evaluate2) edge node [below,sloped] { $G_{\rm Tele}=f(T,T_{\rm ax},T_{\rm vec})$} (decide4);
	\path (decide4) edge node [below,sloped] {$ f=(\frac{3}{2}T_{\rm ax}+\frac{2}{3}T_{\rm ten})^2$} (decide5);
	\path (decide4) edge node [sloped,below] {\hspace{-0.5cm}$f=a_1 T_{\rm ax}+a_2 T_{\rm ten}+a_3 T_{\rm vec}$} (decide7);
	\path (decide4) edge node [below,sloped] {$f=f(T)$} (decide6);

		\path (decide7) edge node [sloped,above] {\scriptsize{\hspace{-0.5cm}$a_1=3/2, a_2=2/3$}} (TEGR);
			\path (decide7) edge node [sloped,below] {\scriptsize{$a_3=-2/3$}} (TEGR);
				\path (decide6) edge node [right] {$f=T$} (TEGR);
					\path (stop) edge node [sloped,above] {$G_2=0,\tilde{G}_4=1$} (TEGR);
					\path (stop2) edge node [sloped,below] {$G_2=0,F=1$} (TEGR);
	\path (decide) edge node [right] {$F(\phi)=1$} (stop);
		\path (decide3) edge node [below,sloped] {$F=F(\phi)$} (stop2);
	\path (decide) edge node [above,sloped] {$\tilde{G}_4(\phi)=0$} (stop2);
	\path (identify) edge node [right] {$G_3=G_5=0,G_4=\tilde{G}_4(\phi)$} (expert);
	\path (identify) edge node [below,sloped] {$G_5=0$} (expert2);
	
	\path (identify) edge node [below,sloped] {$G_4=1,G_5=0$} (expert4);
	\path (expert) edge node [below] {\scriptsize{$\tilde{G}_4(\phi)=1$}} (init);
	\path (expert4) edge node [above,sloped] {$G_2=G_3=0$} (GR);
	\path (expert) edge node [above,sloped] {$G_2=0,$} (GR);
		\path (expert) edge node [below,sloped] {$\tilde{G}_4=1$} (GR);
		\path (init) edge node [above,sloped] {$G_2=0$} (GR);
			\path (expert2) edge[bend left=40] node [above,sloped] {$G_2=G_3=0,\,G_4=1$} (GR);
	\end{scope}
	\end{tikzpicture}}
	\caption{Relationship between Teleparallel Horndeski and various theories.}
	\label{fig}
\end{figure}

\newpage

\section*{Acknowledgements}
The authors would like to acknowledge networking support by the COST Action GWverse CA16104. This article is based upon work from CANTATA COST (European Cooperation in Science and Technology) action CA15117, EU Framework Programme Horizon 2020. S.B. is supported by Mobilitas Pluss No. MOBJD423 by the Estonian government. S.B. would like to thank J.L.S. and his group for the hospitality at the Institute of Space Sciences and Astronomy at the University of Malta, during the preparation of this work. The authors would like to thank Manuel Hohmann and Tomi Koivisto for helpful discussions.

\end{document}